\documentclass[showpacs,amssymb,floatfix,twocolumn]{revtex4}[12pt]
\usepackage{graphicx}
\usepackage{dcolumn}
\usepackage{graphics}
\usepackage{epstopdf}
\usepackage{pdfpages}
\usepackage{amssymb}
\usepackage{amsmath}
\usepackage{bm}

\begin{document}
\title{Engineering of the extremely high $Q$ factor
in two subwavelength dielectric resonators}
\author{E.N. Bulgakov, K.N. Pichugin, and A.F. Sadreev}
\affiliation{$^1$Kirensky Institute of Physics Federal Research
Center KSC SB RAS 660036 Krasnoyarsk Russia\\}
\begin{abstract}
The high-Q  {\it subwavelength} resonances in an isolated
dielectric disk  modes can  be achieved by avoided crossing
(anticrossing) of the nonorthogonal TE resonances under variation
of the aspect ratio as it was reported by Rybin {\it et al}
\cite{Rybin2017}. Traversing over two parameters, the aspect ratio
and the distance between two disks, enhances the $Q$ factor by
several times compared to the case of the former case of
one-parametric avoided crossing in the isolated disk. Therefore
successive  two-parametric avoided crossing gives multiplicative
gain in the $Q$-factor as it was expected. However if for the
single disk its orthogonal resonant modes do not undergo avoided
crossing, a presence of the second disk gives removes this
restriction and gives rise to the avoided crossing of these modes
that enhances the $Q$ factor by two orders in magnitude compared
to the case of single disk. Respectively the multipolar
decomposition of the anti-bonding resonant mode demonstrates
conversion from lower to higher orders of the multipole modes
similar to that as shown by Chen {\it et al} \cite{Chen2019}.
These results are interpreted by that the field configuration at
the maximal conversion becomes close to the Mie resonant mode with
high orbital momentum in equivalent  sphere. For $m=1$ the
resonant modes leakage into both type continua that substantially
lowers the Q-factor of the subwavelength resonant modes.

\end{abstract}
 \maketitle

\section{Introduction}
High index dielectric nanoresonators support different types of resonant modes
which have finite $Q$ factor.
However, the problem is to achieve high-Q factor  in  a  single
 or at least two {\it subwavelength}  dielectric  nanoresonators because of
leakage of the resonant modes into the radiation continuum.
In general, there is a compromise between high Q factors and small
mode volumes due to the fact that larger resonators are required
to increase round-trip travel time for Q-factor enhancement, as is
the case for whispering gallery modes
\cite{Braginsky1989,Cao2015}. Thus it is rather challenging for
optical resonators to support resonances of simultaneous sub
wavelength mode volumes and high Q factors.
The traditional way for increasing of the $Q$  factor of optical
cavities is a suppression of coupling of the resonant modes with  the
radiation continua of either free space or with open channels of
photonic crystal waveguides. The well known examples of such a strategy
are the  Fabry-Perot resonator or whispering gallery modes  for which the Q
factor reaches extremal values.

The decisive breakthrough came with paper by Friedrich and Wintgen
\cite{Friedrich1985} which put forward the concept of destructive
interference of different resonant modes leaking into the
continuum. When two resonances avoid each other as a function of a
certain continuous parameter, interferences of the resonant states
cause the width of one of the resonance states may vanish exactly.
Since it remains above threshold for decay into the continuum,
this state becomes a bound state in the continuum (BIC) although
each resonant state has a finite width.
However, these BICs exist provided that they embedded into a
single continuum of propagating modes of a directional waveguide.
In photonics the optical BICs embedded into the radiation
continuum specified by continuous spectra $\omega=ck$ (the light
line) can be realized by many ways. The first way is realized in
an optical cavity coupled with a few continua of 2d photonic
crystal (PhC) directional waveguide \cite{BS2008}. Alternative way
is the use of periodic PhC systems (gratings) or arrays of
dielectric particles in which resonant modes leak into a
restricted number of diffraction continua
\cite{Shipman2005,Marinica,Hsu2013,PRA2014,PRA96}. Although the
BICs can exist only in infinite periodical arrays because of  the
non-existence theorem \cite{Colton,Silveirinha14}, finite arrays
demonstrate resonant modes with extremely high $Q$ factor
(quasi-BICs) which grows quadratically, cubically or even
exponentially \cite{Bulgakov2018b,Bulgakov2019a} with the number
of particles. Even arrays of five dielectric particles demonstrate
the $Q$ factor exceeding the $Q$ factor of individual particle by
three orders in magnitude \cite{Taghizadeh2017}.

Although individual dielectric resonator can not support BIC
the concept of avoided crossing of resonances is turned out fruitful even in
isolated subwavelength high-index dielectric resonators that allowed
 to achieve high $Q$ resonant modes (super cavity modes)
\cite{Rybin2017,Bogdanov2019}.  Such super cavity modes
originate from hybridization of the resonant modes,
specifically the Mie-type resonant mode and the Fabry-P\'{e}rot
resonant mode under variation of the aspect ratio of the
dielectric disk. As a result  a significant enhancement of
the $Q$ factor  by one order in magnitude was achieved \cite{Rybin2017}.
It is worthy to notice  the
formation of long-lived resonances near avoided crossings was
reported for traversing the distance between quantum dots \cite{Rotter2005} and
for deformation of optical microcavities \cite{Wiersig2006}. By
one order  in magnitude the $Q$ factor enhancement was predicted in Refs.
\cite{Boriskina2006,Boriskina2007,Song2010,Benyoucef2011} for avoided crossing of
high-lying whispering gallery modes for variation of distance
between dielectric cylinders.

\begin{figure}[ht!]
 \centering
\includegraphics[width=0.75\linewidth]{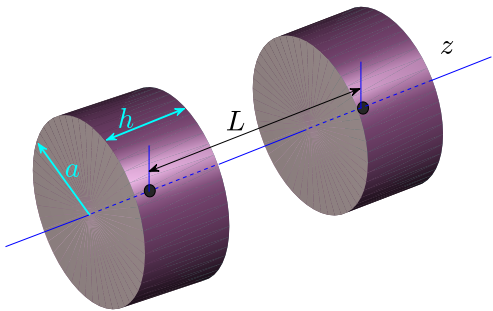}
\caption{Two coaxial dielectric disks with $\epsilon=12$ separated
by distance $L$ measured between the centers of disks.}
\label{fig1}
\end{figure}

In this paper we consider  two identical subwavelength dielectric
coaxial silicon disks with the permittivity $\epsilon=12$ and negligible
material losses at the wavelength $\lambda=1.5\mu m$ \cite{Li1980}. The aim of
consideration is to show ultimate enhancement of the $Q$ factor
under variation of  two parameters: the aspect ratio and the
distance between the disks  as sketched in Fig. \ref{fig1}. For such a
simplest engineering of two disks numerous events of avoided crossing of
the TE or TM resonances of low order occur in the vicinity of which the
hybridized resonant modes of two disks becomes close to the high order Mie resonant mode
of dielectric sphere. Along with
fundamental interest in the ways to enhance the $Q$
factor of the subwavelength dielectric resonators there is also a
motivation caused by current experimental and technological
facilities in fabrication of dielectric disks
\cite{Sadrieva2019,Chen2019}. It is preferable to traverse over
the distance between two disks that varies the effective height of disk dimer
compared to the height of the isolated disk.
In contrast to Refs. \cite{Boriskina2007,Benyoucef2011,Wang2019}
the azimuthal index $m$ is preserved in the system of two coaxial
disks that allows to consider resonances specified by $m$. In the
present letter we focus on the case $m=0$ in which the solutions
are separated by polarization with $H_z=0$ (TE modes) and $E_z=0$
(TM modes) and the case $m=1$. In the last case the resonant modes leak into both
E and H channels that makes the case $m=0$ more favorable with
respect to the $Q$ factor. Our calculations
indeed show that a gain in the Q-factor in the sector $m=1$ yields several times
to the case $m=0$.
\section{Avoided crossing of resonances of single dielectric disk
for traversing over aspect ratio}
In general the resonant modes and their eigenfrequencies are given by
solving the time-harmonic source-free Maxwell's equations \cite{Joan,Lalanne2018}
\begin{equation}\label{ME}
    \left(%
    \begin{array}{cc}
  0 & -\frac{i}{\epsilon}\nabla\times \\
  i\nabla\times & 0\\
\end{array}%
\right)
\left(%
\begin{array}{c}
  \mathbf{E}_n \\
  \mathbf{H}_n \\
\end{array}%
\right)=k_n\left(%
\begin{array}{c}
  \mathbf{E}_n \\
  \mathbf{H}_n \\
\end{array}%
\right)
\end{equation}
where $\mathbf{E}_n$ and $\mathbf{H}_n$ are the EM field
components defined in Ref. \cite{Lalanne2018} as quasi normal modes
which are also known as resonant states
\cite{More1973,Muljarov2010} or leaky modes \cite{Snyder1984}. It
is important that they can be normalized and the orthogonality
relation can be fulfilled by the use of perfectly matched layers
(PMLs) \cite{Lalanne2018}. With the exception very restricted
number of symmetrical particles (cylinders, spheres) Eq.
(\ref{ME}) can be solved only numerically
 in particular by COMSOL Multiphysics.
The eigenfrequencies of resonances are complex
$k_n=\omega_n+i\gamma_n$. In what follows the light velocity is
taken unit.

\begin{figure}[ht!]
\centering
\includegraphics[width=0.9\linewidth]{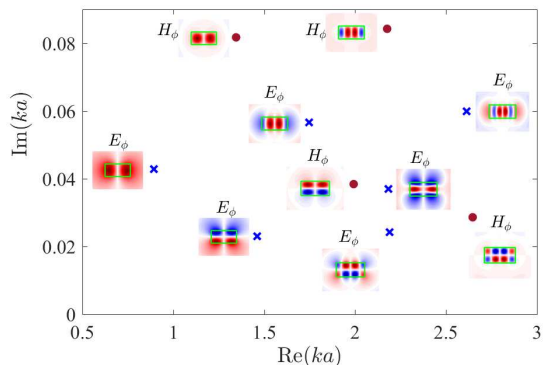}
\caption{The resonant eigenfrequencies (closed circles) and
corresponding resonant modes of dielectric disk with  $h=a$ and
permittivity $\epsilon=12$ in silicon at $\lambda=1.5\mu m$. The
TE resonances are marked by crosses and the TM resonances are marked
by closed circles.}
\label{fig2}
\end{figure}

In what follows we consider the sector with zero azimuthal index
$m=0$ in which all components of EM field are independent of the
azimuthal angle $\phi$. Then it follows from the Maxwell equations
(\ref{ME}) that for the TE polarization ($E_z=0$) we have three
nonzero components of EM field $E_{\phi}(r,z), H_r(r,z),
H_z(r,z)$. Respectively for the TM polarization with $m=0$ we have
the nonzero components $H_{\phi}(r,z), E_r(r,z), E_z(r,z)$. The
components $E_{\phi}$ and their complex resonant frequencies for
the TE modes are shown in Fig. \ref{fig2} by crosses. Respectively
the components $H_{\phi}$ and their frequencies are shown in Fig.
\ref{fig2} by closed circles. It is worthy to lay stress that we
consider only those resonant modes whose characteristic wave
length does not exceed the disk's diameter $2a$ and thickness $h$,
so called the subwavelength resonant modes with resonant
frequencies $\omega_n \leq 2\pi/a$ and $\omega_n \leq 2\pi/h$.

{\bf TE resonances}\\

\begin{figure}[ht!]
\centering
\includegraphics[width=0.9\linewidth]{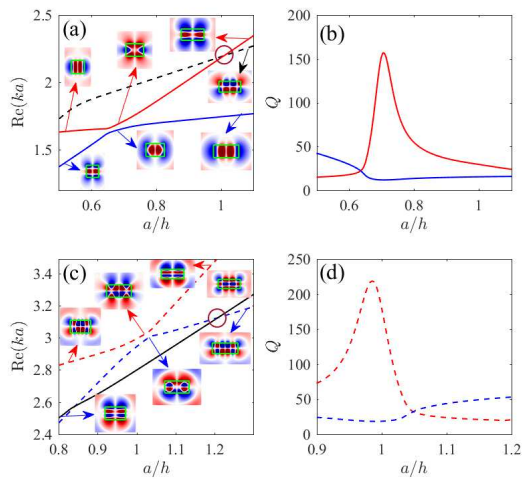}
\caption{(a) Avoided crossing of two TE resonances whose modes are
symmetric relative to $z\rightarrow -z$ and their $Q$ factors
versus the aspect ratio $a/h$ in isolated silicon disk. Insets
show the profiles of tangential component of electric field
$E_{\phi}$. (b) The behavior of the $Q$ factor of the
corresponding resonances. (c) Avoided crossing of two
anti-symmetric (anti-bonding) TE resonances and (d) corresponding
behavior of the $Q$ factor.} \label{fig3}
\end{figure}

 Fig. \ref{fig3} (a) shows the evolution of complex resonant
eigenfrequencies in traversing over the aspect ratio of the
isolated silicon disk with the permittivity $\epsilon=12$. Because
of the symmetry of the disk relative to inversion of the disk's
axis only the resonances of the same symmetry undergo the avoided
crossing while the resonances of the opposite symmetry cross each
other as highlighted by circles. Similar to Refs.
\cite{Rybin2017,Chen2019} the $Q$ factor reaches a magnitude
around 150 (Fig. \ref{fig3} (b)) in the vicinity of the avoided
crossing showing an enhancement by one order in magnitude. That is
typical for subwavelength dielectric resonators \cite{Rybin2017}.
We present also another case of the avoided crossing in Fig.
\ref{fig3} (c) which results in the $Q$ factor around 220 as shown
in Fig. \ref{fig3} (d). Although this case goes beyond the
subwavelength limit the case is interesting by the avoided
crossing of antisymmetric resonances.

\begin{figure}[ht!]
\centering
\includegraphics[width=0.9\linewidth]{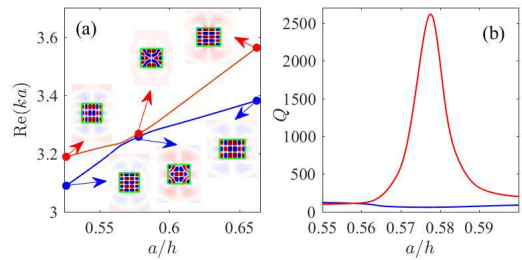}
\caption{(a) Avoided crossing of the TM-resonances in isolated disk.
Insets show the profiles of the component of magnetic field $H_{\phi}$.
(b) The behavior of the $Q$ factors.}
\label{fig4}
\end{figure}

{\bf TM resonances}\\
The case of TM resonances demonstrates different scenario for the
avoided crossing shown in Refs. \cite{Heiss2000,Rotter2005}.
Typically for the avoided crossing we observe repulsion of real
and imaginary parts of resonances both as it was shown above for
the TE polarization for variation of the aspect ratio. For
traversing over the aspect ratio one can observe from Figs.
\ref{fig3}   that the resonant modes are interchanging with full
hybridization at the minimal distance between their frequencies
where the $Q$ factor reaches maximum. The case of the TM
resonances is more sophisticated. One can see from Fig. \ref{fig4}
(a) that the real parts of resonant frequencies stick together in
the range of the aspect ratio $a/h$ from 0.57 till 0.59 while the
imaginary parts demonstrate strong repulsion
with extremal enhancement of the $Q$ factor as shown in Fig. \ref{fig4} (b).

\begin{figure}[ht!]
\centering
\includegraphics[width=0.9\linewidth]{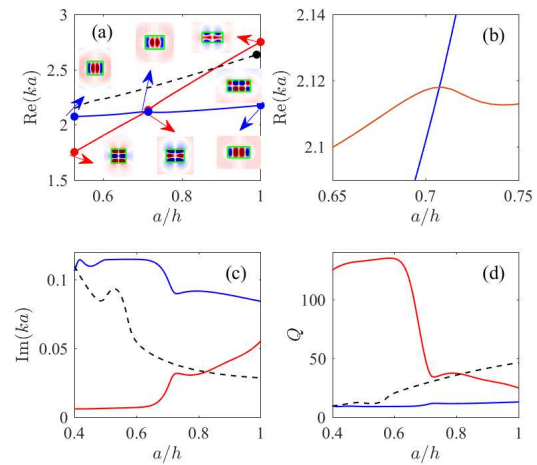}
\caption{Crossing of two TM-resonances in isolated disk. Behavior
of real (a), (b) and imaginary (c) parts of complex resonant
frequencies of two anti-bonding resonances (solid lines) where
insets show the profiles of component $H_{\phi}$ magnetic field.
(d) The behavior of the $Q$ factors. The bonding resonance is
shown by dash line.} \label{fig5}
\end{figure}

Moreover Fig. \ref{fig5} (a) demonstrates that the real parts of the TM resonances
are crossing. Also one can see that the resonant modes are not
interchanging for traversing over the aspect ratio in contrast to former cases
shown in Figs. \ref{fig3} and \ref{fig4}.
One or another behavior of resonances for parametric traversing
depends on difference between the couplings of the resonant modes
with the radiation continuum, i.e., on the imaginary parts of the
resonances. When the difference between couplings becomes smaller
we observe the avoided crossing of resonances but when the
coupling are strongly different we observe the crossing of
resonances \cite{Heiss2000,Rotter2005}.

\section{Two-parametric avoided crossing}
\begin{figure}[ht!]
\includegraphics[width=1\linewidth]{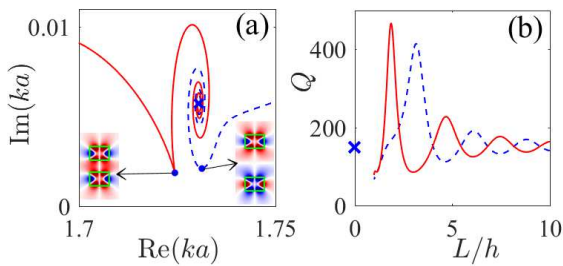}
\caption{(a) Evolution of the hybridized bonding (solid) and
anti-bonding (dash) resonances
 under variation of the distance between the disks $L$ at the aspect ratio $a/h=0.7$.
Insets show the tangential component $E_{\phi}$ of electric field.
Crosses mark the resonances of the isolated disk which become
degenerate at large distances. (b) Evolution of  the $Q$ factors
vs the distance.} \label{fig3fromPRB}
\end{figure}

\begin{figure}[ht!]
\centering
\includegraphics[width=0.9\linewidth]{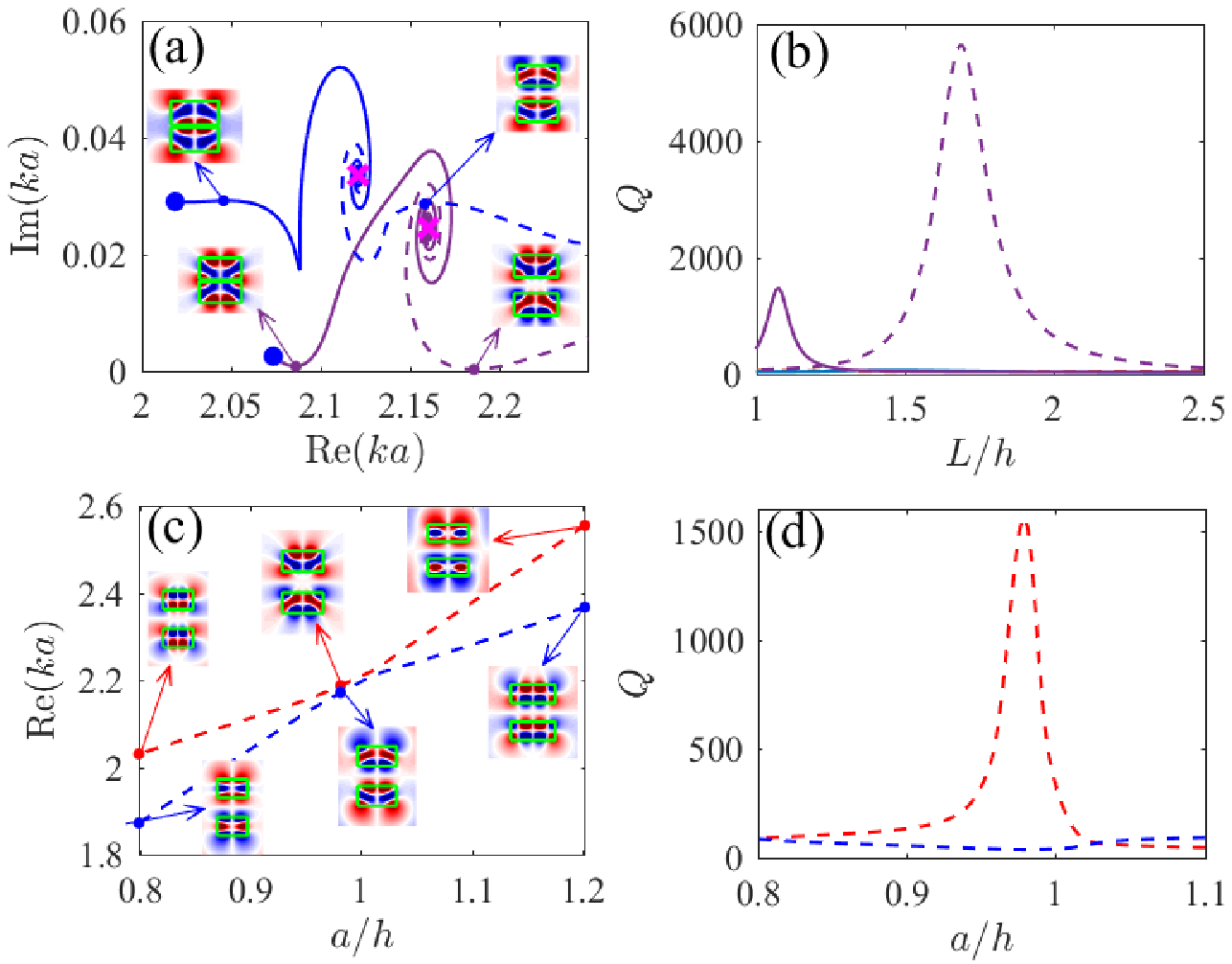}
\includegraphics[width=0.6\linewidth]{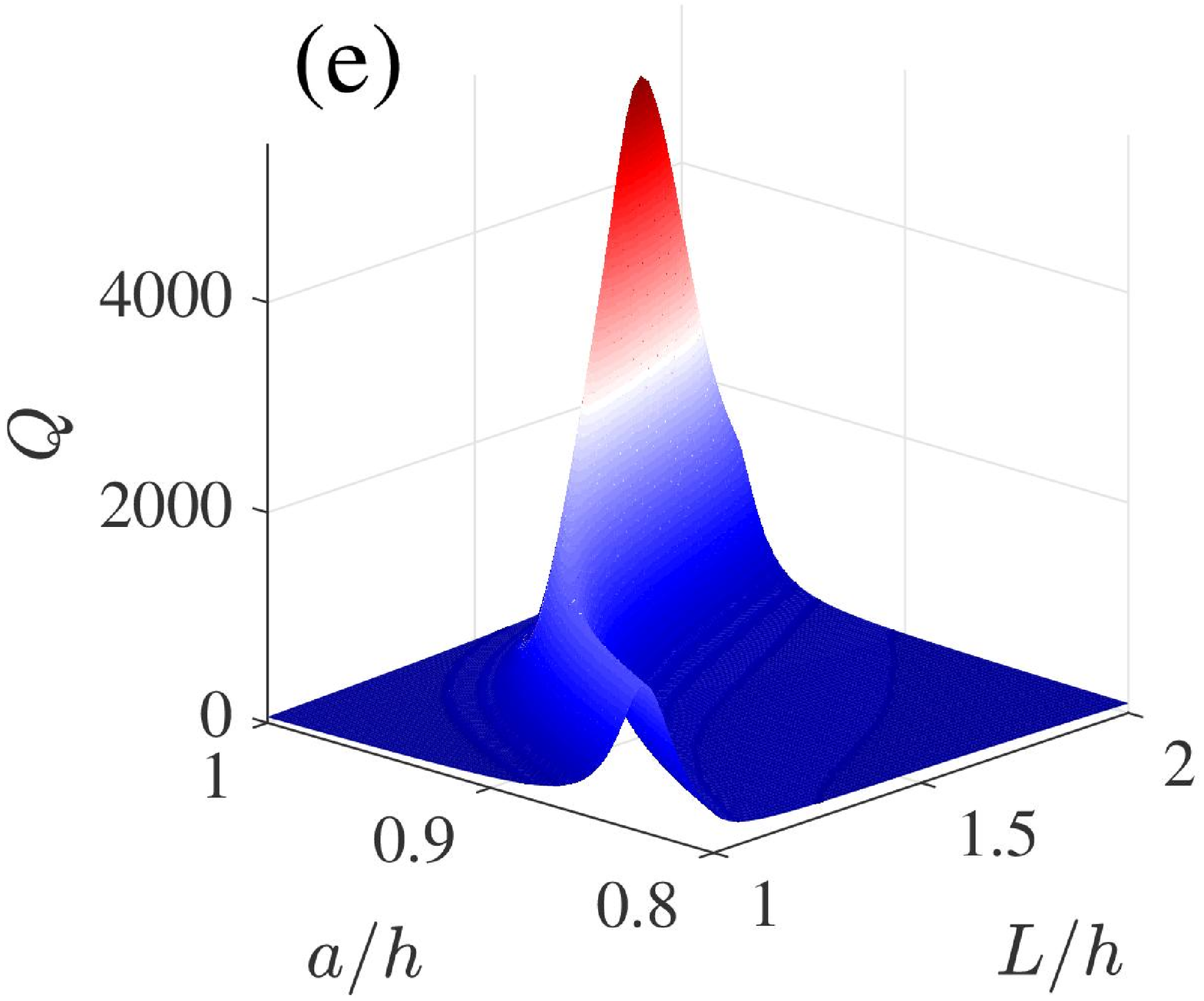}
\caption{(a) Avoided crossing of TE resonances originated from the
resonances of isolated disk marked by crosses and (b) their $Q$
factors vs the distance $L/h$ at the aspect ratio $a/h=0.96$.
Solid and dash lines refer to the bonding  and anti-bonding
resonant modes of separate disks. Closed circles mark the case
$L=h$ when two disks stick each other with double height $2h$. (c)
Avoided crossing of two resonances whose modes are antisymmetric
relative to $z\rightarrow -z$ and (d) their $Q$ factors vs aspect
ratio $a/h$ in the system of two disks separated by distance
$L=2.07a$. Insets show profiles of electric field $E_{\phi}$. (e)
he total picture of the $Q$ factor vs aspect ratio and distance
between disks.}
 \label{fig6}
\end{figure}

In the system of  two dielectric disks we have a second parameter
to vary, the distance between the disks, $L$. This distance is
measured between L is measured between volumetric centers as shown
in Fig. \ref{fig1}, therefore minimal distance is $h$.  First of
all variation $L$ provides technological advance over variation of
the aspect ratio of isolated disk because of continual variation
of effective height of disk's dimer equal $h+L$. However a
presence of the second disk brings on two new effects. The first
is an interaction between the disks. The radiation of the first
disk is scattered by the second disk resulting in coupling between
the disks that lifts the degeneracy of the resonances of two disks
with further avoided crossing of resonances with enlargement of
the distance between the disks as shown in Figs. \ref{fig3fromPRB}
and \ref{fig6}(a). These results are reported in detail in our
former publication \cite{Pichugin2019}. In order to remain in the
subwavelength regime for two disks we keep inequality
$k<\frac{2\pi}{h+L}$. For reader's convenience we present one of
scenarios of the behavior of resonances for variation of distance
between disks provided that each disk is optimized to have maximal
$Q$-factor with the respect to aspect ratio $a/h=0.7$
\cite{Rybin2017} in Fig. \ref{fig3fromPRB}. When the distance
between the disks is large enough the resonances can be considered
as degenerate. These limiting cases are marked by crosses in Figs.
\ref{fig3fromPRB} and Fig. \ref{fig6}. We choose those resonant
mode which has the high $Q$-factor around 150 which corresponds to
the "supercavity" mode in terminology by Rybin {\it et al}
\cite{Rybin2017}. It is highlighted by circle in Fig. \ref{fig3}
(a). As disks are coming closer the coupling between the disks
hybridizes the modes by symmetric and antisymmetric ways as shown
in Fig. \ref{fig3fromPRB} and splits the resonances which evolve
by spiral way because of accumulation of phase $\exp(-2ikL)$ in
processes of mutual scattering of leaky resonant modes by disks
\cite{Pichugin2019}. Respectively, real and imaginary parts of
complex resonant frequencies oscillate with $L$ as well as the
$Q$-factor as seen from Fig. \ref{fig3fromPRB} (b) with maximal
value around 450 that exceeds the $Q$-factor of isolated disk by
three times. Therefore two-parametric avoided crossing of two
resonant modes of the same symmetry relative to inversion of the
symmetry axis z gives a multiplicative gain in the Q-factor.

\begin{figure}[ht!]
\centering
\includegraphics[width=1\linewidth]{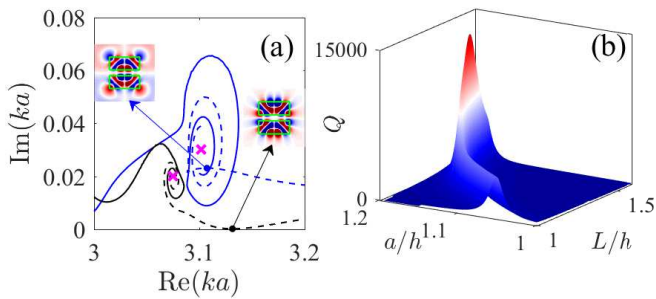}
\caption{(a) Evolution of the higher lying TE resonances in
traversing with the distance between the disks for $a/h=1.17$. (b)
The $Q$ factor vs the distance between the disks $L/h$ and their
aspect ratio $a/h$.} \label{fig7}
\end{figure}

However what is more remarkable, the presence of the second disk
opens new ways for the avoided crossing which were prohibited in
the single disk. For example, as Fig. \ref{fig3} shows the
resonances of opposite symmetry plotted by solid and dash lines do
not notice each other. The presence of the second disk lifts this
symmetry restriction giving rise to a new series of avoided
crossings of resonances. These crossings are highlighted by
circles in Figs. \ref{fig3} (a) and (c).
Fig. \ref{fig6} shows how the avoided crossing of the resonances
with opposite symmetry at the point $a/h=0.96$ results in extremal
enhancement of the $Q$ factor.
It is clear that this enhancement of the $Q$ factor is not unique
because the case of two disks doubles the number of
avoided crossings. For example Fig. \ref{fig3} (c) shows
another case of crossing of the resonances of opposite symmetry
around the aspect ration $a/h=1.2$ of isolated disk.
Fig. \ref{fig7} (a) demonstrates as these resonances
avoid each other with enlargement of the distance between
disks around this aspect ratio. Fig. \ref{fig7} (b)
shows that in order to engineer  extremal values of the $Q$ factor
it is necessary to accurately tune both parameters, the aspect ratio and
the distance between disks. As a result an enhancement reaches 15000 that
exceeds the $Q$ factor of the isolated disk is 68 times.
Although these resonances have more high frequencies around $ka\approx 3.1$,
they are still belong to subwavelength range because of the inequality
$k<2\pi/2h+L\approx\pi/a$.
\begin{figure}[ht!]
\centering
\includegraphics[width=1\linewidth]{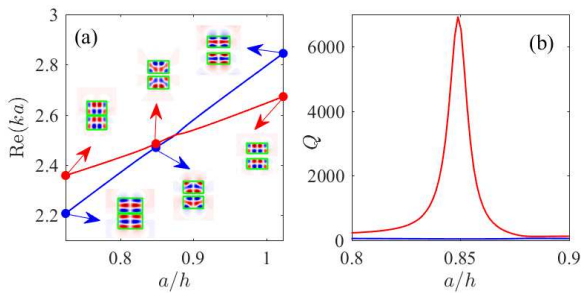}
\caption{Evolution of the TM resonances (a) and $Q$ factor (b) in
traversing with the height of the disks at $L/a=1.48$. Insets
shows profiles of the TM resonant modes $H_{\phi}$.}
 \label{fig8}
\end{figure}
Fig. \ref{fig8} shows as the TM resonances are crossing for
traversing over the distance between the disks and aspect ratio.
Nevertheless the $Q$ factor behaves in the same way as the
resonances would undergo the avoided crossing as it was
demonstrated in previous Figures for the TE resonances.
\section{One disk at metal surface}
\begin{figure}[ht!]
\centering
\includegraphics[width=0.45\linewidth]{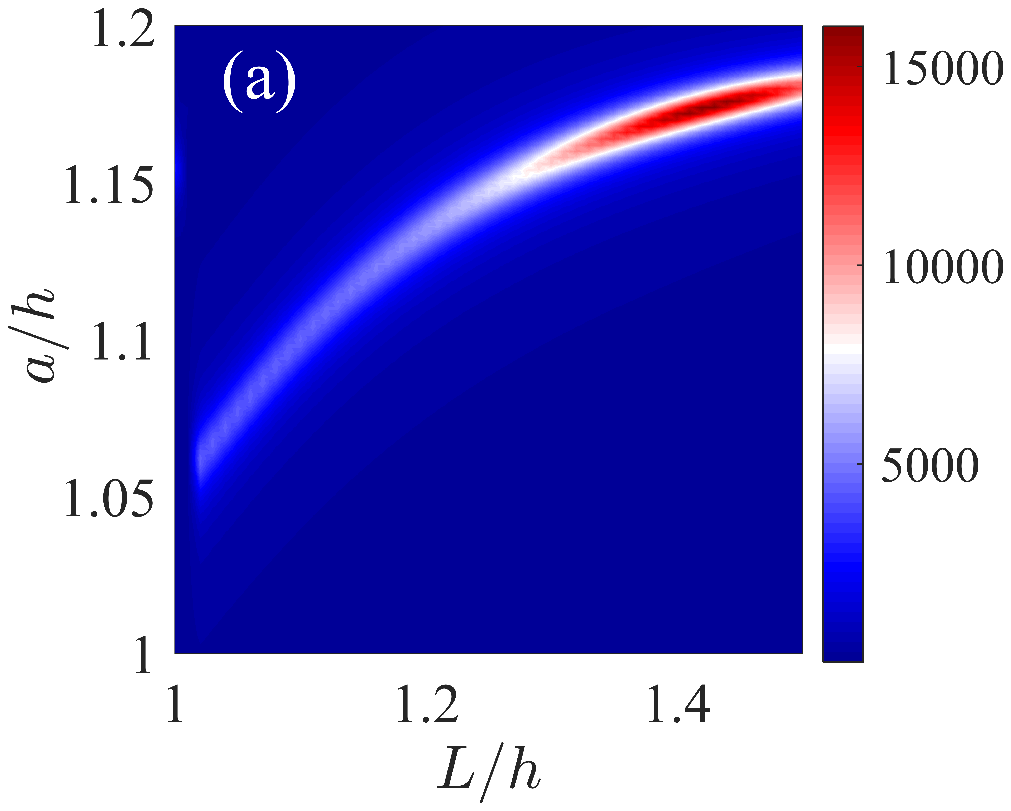}
\includegraphics[width=0.45\linewidth]{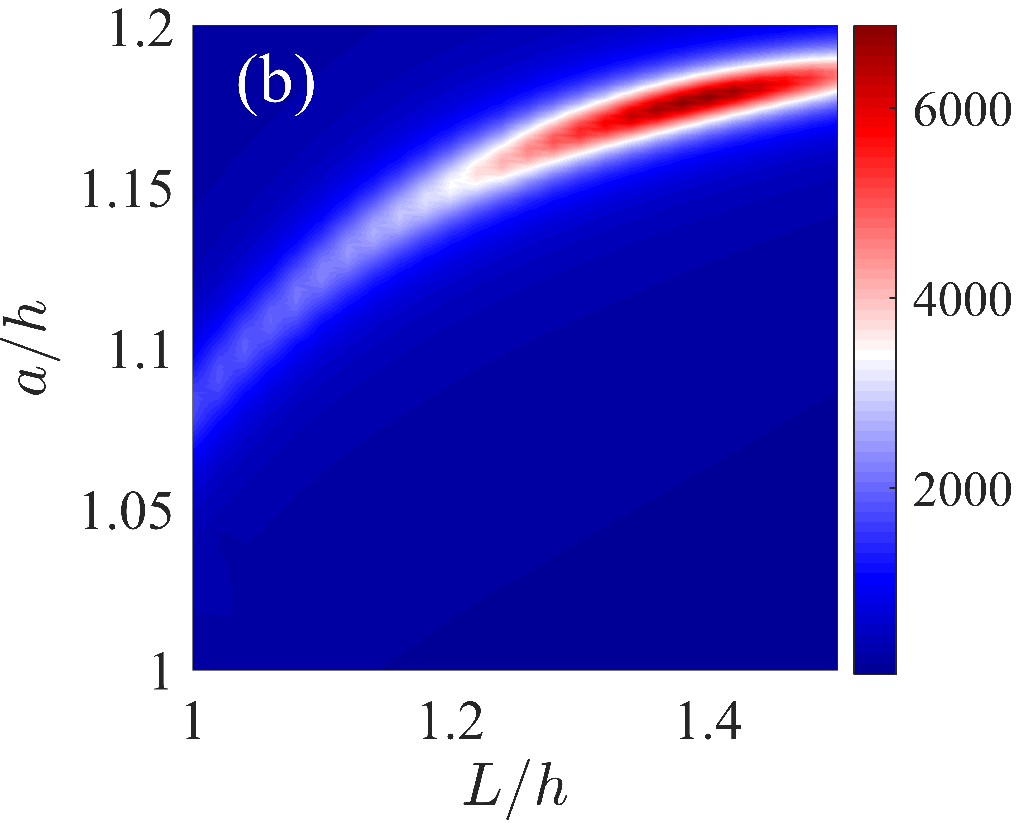}
\caption{$Q$ factor vs the aspect ration and distance (a) between
disks and (b) between disk and its image in metal silver mirror
with refractive index $n=0.14+11i$.} \label{fig13}
\end{figure}

Also one can manage to substitute the system of two coaxial disks
by one disk whose axis is normal to  the surface of metal mirror.
Although the systems are not equivalent because of boundary
conditions at the metal surface (the magnetic field as a pseudo
vector is to be zero at the surface) and ohmic losses.
Nevertheless Fig. \ref{fig13} shows that behavior of the $Q$
factor is very similar to the case of two dielectric disks. Here
$L$ is doubled distance between the disk center and metal surface.

\section{Cancellation of the lowest-order terms in a multipole
radiation for avoided crossing} Any solution of the Maxwell
equations can be expanded in terms of electric and magnetic
spherical harmonics \cite{Jackson}
\begin{equation}\label{MN}
\mathbf{E}(\mathbf{x})=\sum_{l=1}^{\infty}\sum_{m=-l}^l[a_{lm}\mathbf{M}_{lm}+
b_{lm}\mathbf{N}_{lm}].
\end{equation}
Then the relative radiated power of each electric and magnetic
multipoles of order $l$ is given by squared amplitudes of
expansion \cite{Jackson}
\begin{equation}\label{Plm}
    P_{lm}=P_{lm}^{TE}+P_{lm}^{TM}=P_0^{-1}[|a_{lm}|^2+|b_{lm}|^2]
\end{equation}
where $P_0$ is the total power radiating
through the sphere with large radius
\begin{equation}\label{PW}
    P_0={\sum_{l=1}^{\infty}\sum_{m=-l}^l[|a_{lm}|^2+|b_{lm}|^2]}.
\end{equation}
It is intuitively clear that the sharp enhancement
of the $Q$ factor is result of cancellation of the lowest-order terms in a multipolar
expansion of the far-field radiation, distinct from the
near-field multipole symmetry \cite{Johnson2001,Chen2019}.

\begin{figure}[ht!]
\centering
\includegraphics[width=0.8\linewidth]{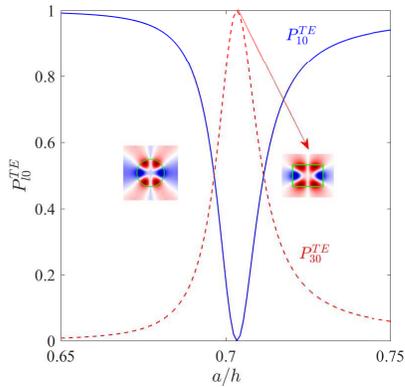}
\caption{Evolution of the TE multipole radiated powers from one disk vs the aspect ratio in
isolated disk for the case shown in Fig. \ref{fig3}.}
 \label{fig9}
\end{figure}

For the case of avoided crossing with traversing of the aspect
ratio of isolated disk Chen {\it et al} has interpreted that
significant Q-factor enhancement at is attributed to strong
redistribution of radiation that originates from suppression of
electric dipole  radiation as shown in Fig. \ref{fig9}. The case
corresponds to the avoided crossing of resonant modes shown in
Fig. \ref{fig3} (a).
Insets in Fig. \ref{fig9} clearly demonstrate that the field configuration at the
maximal conversion is very close to the ideal octuple resonant mode in the
dielectric sphere with the radius given by the equality of volumes of the sphere
and disk with the aspect ratio $a/h=0.706$: $4\pi R^3/3=\pi a^2h$. For equal
volume, permittivity and frequency $kR=1.63$ sphere has an advantage
in the $Q$ factor $Q=194$ compared to the disk with $Q=157$. As it follows
below this advantage will preserve also for higher resonances and is intuitively
clear because of less area of sphere compared to the disk.

\begin{figure}[ht!]
\centering
\includegraphics[width=0.8\linewidth]{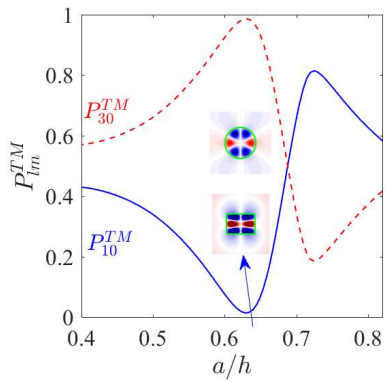}
\caption{Evolution of the multipolar TM powers from disk vs the aspect ratio
 for the case shown in Fig. \ref{fig7}.}
\label{fig10}
\end{figure}

Fig. \ref{fig10} demonstrates that in spite of different behavior of the TM resonances
shown in Fig. \ref{fig5} the radiated multipolar powers also undergoes
similar conversion as shown in Fig. \ref{fig9}.
The corresponding Mie octuple TM resonant mode of the sphere (the component $H_{\phi}$
with frequency $kR=1.94$ and $Q=194$ is similar to the TM hybridized mode of the disk
with frequency $ka=2.12$ and $Q=135$ as shown in Fig. \ref{fig10}.

\begin{figure}[ht!]
\centering
\includegraphics[width=0.8\linewidth]{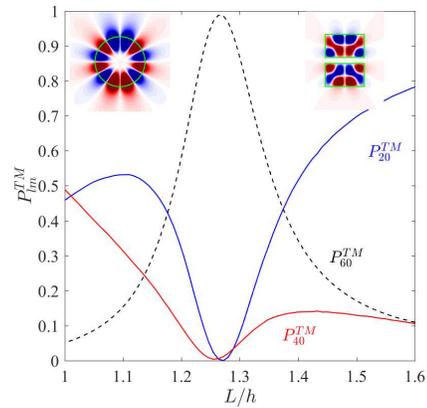}
\caption{Evolution of the radiated power from two disks vs the distance
between disks for the case shown in Fig. \ref{fig6} ($h=1.038a$).}
 \label{fig11}
\end{figure}

Next, let us consider the case of two disks. We start with the case presented
in Fig. \ref{fig6} (a) and (b) when for avoided
crossing of the TE resonances of opposite symmetry the $Q$ factor approaches 5500.
Corresponding evolution of the multipolar powers $P_{l0}^{TE}$ is presented in Fig.
\ref{fig11}.
Similar to the isolated disk we introduce equivalent sphere by
equality of the volume of  disk's dimer and sphere  $4\pi
R^3/3=\pi a^2(h+L)$. The $Q$ factor of two disks 5500 for the
hybridized resonance $ka=2.2$ substantially yields the $Q$ factor
23100 of the Mie resonant mode $l=4, m=0$ with frequency $kR=2.68$
of the equivalent sphere shown in left inset of Fig. \ref{fig11}.
In spite of that one see that the hybridized anti-bonding resonant
mode of two diks shown in right inset of Fig. \ref{fig11} has the
same morphology that explains so high $Q$ factor of two disks.
While for the avoided crossing of TE resonances of the same
symmetry shown in Fig. \ref{fig3fromPRB} the hybridized bonding
resonant modes have the morphology cardinally different from the
Mie resonant modes of the sphere. As result the bonding resonances
have no extremely high $Q$ factors.

Figs. \ref{fig9}--\ref{fig11} evident that the sharp $Q$
factor enhancement is intrinsically connected with multipolar conversions from lower
to higher orders.  Destructive interference between two resonances
underlie these phenomena when the resonances undergo avoided crossing.
For that process the system supports the hybridized modes which
becomes maximally close to the Mie resonant modes of the
dielectric sphere with the high orbital momentum index $l=4$.
Alongside this the system of two disks always yields to isolated sphere
of the same volume in the $Q$ factor since the sphere's surface has less area than the isolated disk or
the system of two disks. Respectively the sphere radiates less power compared to
any dielectric resonator of the same volume as the volume of sphere.

\section{Sector $m=1$}
\begin{figure}[ht!]
\centering
\includegraphics[width=1\linewidth]{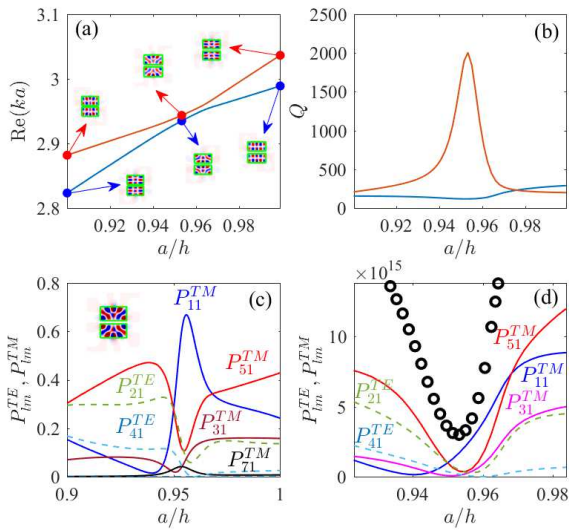}
\caption{Avoided crossing of two resonances (a) and the
$Q$ factor (b) versus the aspect ratio $a/h$ at $L=1.3a$ in the sector $m=1$.
Evolution of the radiated multipolar powers normalized by (c) total power
and (d) by energy inside the disks.}
\label{fig12}
\end{figure}

Above the engineering of high-$Q$ resonances of the disks was considered in
the sector $m=0$ in which the resonant modes radiate into the
TE and TM continua selectively. In this sector  one could expect that
the $Q$ factors of the resonant modes would yield the sector $m=0$.
Fig. \ref{fig12} (a) and (b) shows the evolution of resonances with the
aspect ratio $a/h$ and respectively the behavior of the $Q$ factor in the sector $m=1$
at the distance $L=1.3a$ tuned to the
maximal quality factor. Comparison to the case $m=0$ in Fig. \ref{fig3}
indeed shows that radiation into both continua substantially enlarges radiation
losses of two disks if to compare to the case of $m=0$ shown in Fig. \ref{fig7}.
Fig. \ref{fig12} (c) with plots of the multipolar powers reveals
that avoided crossing with traversing over two parameters is not enough to
cancel simultaneously electric quadruple  and magnetic dipole radiated powers.
However if to normalize the radiated power through the total energy accumulated inside
the disks we observe the resonant dip in the power as shown in Fig. \ref{fig12} (d)
by open circles.
\section{Conclusion and Outlook}
When two resonances of the open resonator are traversed over some
parameter they undergo the avoided crossing (anti crossing) with
hybridization of the resonant modes. Typically the real and
imaginary parts of resonances, both, undergo repulsion up to that
one of the hybridized modes acquires the imaginary part
significantly less compared to the case far from the region of the
avoided crossing. That property of the avoided crossing was
numerously used to achieve high $Q$ factor by traversing of the
aspect ratio of the isolated dielectric resonator \cite{Rybin2017}
or two resonators by traversing of the distance between them
\cite{Wiersig2006,Boriskina2007,Song2010,Benyoucef2011}. In the
present paper we studied the avoided crossings  of resonances and
respectively the $Q$ factor traversing over both parameters, the
aspect ratio of disks and distance between them. One could expect
that the maximal enhancement of $Q$ factor will be  multiplication
of gains achieved by traversing over each parameter independently.
Indeed, if, first,  to vary over aspect ratio in the aim to
achieve maximal Q factor we obtain $a/h=0.7$ as it was reported by
Rybin et al \cite{Rybin2017}. Then variation over the distance
between disks results successive gain in Q factor as shown in Fig.
\ref{fig3fromPRB} that was reported in our previous paper
\cite{Pichugin2019}. In the present manuscript we report results
of independent and full scale variation over two parameters to
reveal unprecedented gains of the $Q$ factor with peaks exceeding
at least by one order in magnitude compared to previous cases in
Ref. \cite{Pichugin2019}.

These peaks are originated from resonant modes which were
orthogonal in single disk and therefore could not contribute into
the avoided crossing for variation of aspect ratio.  Events of
crossing of these modes are highlighted by circles in Fig.
\ref{fig3}. One can see that these points rather far from former
point $a/h=0.7$. The presence of a second disk removes former
symmetrical prohibition to give rise to new series of the avoided
crossings shown in Figs. \ref{fig6} and \ref{fig7}. What is the
important only these avoided crossings result in anti-bonding
resonant modes which are very close to the Mie resonant modes with
high orbital moment of effective sphere as Fig. \ref{fig11}
demonstrates. That explains extremely high Q factor of these
anti-bonding resonant modes. The resulting $Q$ factor exceeds the
$Q$ factor of the isolated disk two orders in magnitude.

The frequencies of the TM resonances undergo a crossing as shown
in Fig. \ref{fig8}. A general morphology of behavior of the
complex frequencies of resonances in general two-level formalism
was established by Heiss \cite{Heiss2000}. Irrespectively for both
types of the avoided crossing we have hybridization of low lying
resonant modes of disks which resembles the higher lying Mie
resonant modes of dielectric sphere with volume equal the volume
of disk's dimer. That results in strong suppression of radiation
from the disk's dimer. The avoided crossing of TE resonances in
the strategy of successive enhancement of the $Q$ factor, at
first, over the aspect ratio, then over the distance shown in
Figs. \ref{fig3} (a) and \ref{fig3fromPRB} give the hybridized
resonant modes cardinally different from the Mie resonant modes of
the sphere.

There is useful tool to understand of a nature of the extremely
high quality factor for the avoided crossing through multipolar
expansions \cite{Jackson}. That tool shed light on the origin of
the high $Q$ factor in the isolated disk
\cite{Chen2019,Bogdanov2019} and the origin of BICs
\cite{Sadrieva2019a}. In the present case of two disks we also
observe that extremal Q-factor enhancement is attributed to strong
redistribution of radiation that originates from multipolar
conversions from lower to higher orders for the case of zero
azimuthal index $m=0$. However for $m=1$ disks radiate in both
continua, TE and TM. As a result such a redistribution turns out
not sufficient to strongly suppress radiation and enhancement of
the $Q$ factor is not so impressive.

We lay stress that these results refer to low lying subwavelength
resonances which are important for numerous applications. Thus
such a tuned dimer successfully meets  the request of high
sensitivity and selectivity sensors which can be integrated in
microsystems \cite{Romano2018}. Because the unprecedent
enhancement of the $Q$ factor is the result of lifting of symmetry
restrictions in the system of two coaxial disks  we expect similar
phenomenon for the disks of other materials with different
permittivity with simple scaling of the $Q$ factor in the form
$Q\sim \epsilon^\alpha$ where for particular case of isolated disk
$\alpha$ was estimated as $\alpha=3.2$ according \cite{Rybin2017}.

It is clear that the phenomenon of the avoided crossing and
respective enhancement of the $Q$ factor would occur with
particles of arbitrary shape when the distance between them  is
varied. The case of two coaxial disks simplifies computations
because the solutions with different azimuthal index $m$ are
independent. Despite that the introduction of a second disk
extends the size of the dielectric resonator more than two times,
it provides more easy technological way to vary a parameter
traversing over which gives rise to the avoided crossing. That
strategy resulted in unprecedent enhancement of the $Q$ factor by
500 times compared to the case of an isolated disk with a given
aspect ratio.

{\bf Acknowledgments}:
The author thanks D.N. Maksimov and Yi Xu for discussions.
This work was supported by RFBR grant 19-02-00055.

\bibliography{arxiv}
\end{document}